\documentstyle[prl,aps,epsf,floats]{revtex}

\begin{document}

%
\twocolumn[\hsize\textwidth\columnwidth\hsize\csname @twocolumn\endcsname
\vskip2pc] 
\narrowtext {\flushleft{\bf Comment on ``Absence of a Slater
Transition in the Two-Dimensional Hubbard Model''}} \newline

While we agree with the numerical results of Ref.\cite{Moukouri:2001}, we
arrive at different conclusions: The apparent opening of a gap at
finite-temperature in the two-dimensional {\it weak-coupling }Hubbard model%
{\it \ }at half-filling.does not necessitate an infinite correlation length $%
\xi $ (Slater mechanism) nor a thermodynamic {\it finite-temperature} metal
insulator transition (MIT). The pseudogap is a crossover phenomenon due to
critical fluctuations in two dimensions, namely to the effect of a $\left(
\pi ,\pi \right) $ spin-density-wave (SDW) $\xi $ that is large compared
with the thermal length.

We use the units of Ref.\cite{Moukouri:2001}. The inset of Fig.1 shows $%
\left\langle n_{\uparrow }n_{\downarrow }\right\rangle $ obtained in Ref.%
\cite{Moukouri:2001} for $N_{c}=36,$ $U=1$ and $N_{c}=64,$ $U=0.5$ along
with the corresponding results obtained\cite{VT} from the local moment sum
rule $\left( T/N_{c}\right) \sum_{q}\chi _{sp}\left( q\right)
=1-2\left\langle n_{\uparrow }n_{\downarrow }\right\rangle $, supplemented
with the relations $\chi _{sp}^{-1}\left( q\right) =\chi _{0}(q)^{-1}-\frac{%
U_{sp}}{2}$ and $U_{sp}=U\left\langle n_{\uparrow }n_{\downarrow
}\right\rangle /\left( \left\langle n_{\uparrow }\right\rangle \left\langle
n_{\downarrow }\right\rangle \right) $. Fig.1 also shows the pseudogap in
the density of states $\rho \left( \omega \right) $ obtained from\cite%
{Moukouri:1999} $\Sigma _{\sigma }^{\left( s\right) }(k)=Un_{-\sigma }+\frac{%
U}{8}\frac{T}{N}\sum_{q}\left[ 3U_{sp}\chi _{sp}(q)+U_{ch}\chi _{ch}(q)%
\right] G_{\sigma }^{0}(k+q)$ which includes the effects of both spin $\chi
_{sp}$ and charge $\chi _{ch}$ fluctuations and satisfies $\frac{1}{2}{\rm Tr%
}[\Sigma _{\sigma }^{\left( s\right) }G_{\sigma }^{0}]=U\left\langle
n_{\uparrow }n_{\downarrow }\right\rangle .$The charge fluctuations are
constrained by the sum rule $\left( T/N_{c}\right) \sum_{q}\left( \chi
_{sp}\left( q\right) +\chi _{ch}\left( q\right) \right) =1.$ As temperature
is lowered from $T=1/20$ to $1/22$ and $1/32$, the pseudogap in $\rho \left(
\omega \right) $ quickly deepens. The distance between the two peaks is in
quantitative agreement with Ref.\cite{Moukouri:2001}. In addition, extensive
comparisons with quantum Monte Carlo (QMC) have shown earlier\cite%
{Moukouri:1999,VT} that our approach agrees quantitatively with QMC, and
contains the same finite-size effects. In particular, $\rho \left( \omega
=0\right) $ is smaller in smaller lattices. Hence, while at $T=1/32$ the
criterion\cite{Moukouri:2001}$\rho \left( \omega =0\right) <1\times 10^{-2}$
is satisfied for $N_{c}=64$ and short $\xi \simeq N_{c}^{1/2}$, we still
need to verify that this reflects the behavior of a large (but not infinite)
correlation length in the thermodynamic limit. That is why we verified that $%
\rho \left( \omega =0\right) <1\times 10^{-2}$ for $N_{c}=$ $128^{2}$ as
well. $\rho \left( \omega =0\right) $ in dynamical cluster approximation
(DCA) has the opposite size dependence and satisfies $\rho \left( \omega
=0\right) <1\times 10^{-2}$ for $N_{c}=64$. Note that for size $128^{2}$, $%
\xi $ already reaches $40$ lattice spacings at $T=1/22$. All of the above
results may be understood analytically from the above equations\cite{VT} by
considering the limiting case where the characteristic frequency in the spin
spectral weight $\chi _{sp}^{\prime \prime }$ becomes smaller than
temperature, (renormalized classical regime). The local moment sum rule
prevents a finite-temperature mean-field transition by letting $U_{sp},$ and
hence $\left\langle n_{\uparrow }n_{\downarrow }\right\rangle $, exhibit a
downturn at $T^{\ast }.$ Below that temperature, $\xi $ grows rapidly but it
becomes infinite only at $T=0.$ Similarly, the opening of the pseudogap with
decreasing temperature can be traced\cite{VT}, in $d=2$, to the singular
contribution of $\chi _{sp}\left( q\right) $ to $\Sigma _{\sigma }^{\left(
s\right) }(k)$ when $\xi $ becomes larger than the single-particle thermal
de Broglie wave length $\xi _{th}=v_{F}/T$. Indeed, in that limit, the
single-particle spectral weight $A\left( {\bf k}_{H},\omega \right) $ at hot
spots is given by $-2\Sigma ^{\prime \prime }\left[ (\omega -\Sigma ^{\prime
})^{2}+\Sigma ^{\prime \prime 2}\right] ^{-1}$ with $\omega -\Sigma ^{\prime
}=0$ and $\Sigma ^{\prime \prime }\left( {\bf k}_{H},0\right) \propto \xi
^{3-d}/\xi _{th}.$ Since $\xi /\xi _{th}$ grows exponentially in the $d=2$
renormalized classical regime, $A\left( {\bf k}_{H},\omega \right) $ can
become exponentially small at $\omega =0$ even without a MIT. In the
analytical approach,\cite{VT} the downturn in $\left\langle n_{\uparrow
}n_{\downarrow }\right\rangle $ and the opening of a deep pseudogap are both
unambiguously driven by a rapidly growing $\xi $ in the SDW channel. The
pseudogap itself is not needed to reinforce the downturn in $\left\langle
n_{\uparrow }n_{\downarrow }\right\rangle $. While the situation is more
subtle than that of Slater, the peaks in Fig.1 are precursors of the SDW
insulator that appears at exactly $T=0$ by the Slater mechanism. The peak
separation in frequency (the gap) is larger than $T^{\ast }$ because
Kanamori screening strongly renormalizes $T^{\ast }$ down$.$ 
\begin{figure}[tbp]
\centerline{\epsfxsize 8cm \epsffile{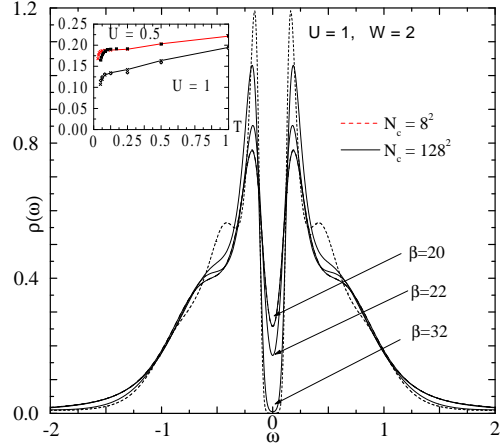}}
\caption{Density of states as a function of $\protect\omega $ for various
lattice sizes, the largest $\protect\beta $ having the largest peak heights.
Inset shows $\left\langle n_{\uparrow }n_{\downarrow }\right\rangle $ with
symbols from Ref.\protect\cite{Moukouri:2001}, except for $\times $ and
lines that represent our calculation.}
\label{fig1}
\end{figure}
Increasing $N_{c}$ in DCA effectively lowers the dimension towards $d=2$,
revealing the effect of $\xi >\xi _{th}$ on $\Sigma $ and $\rho .$

We thank S. Allen, M. Jarrell, P. Lombardo and S. Moukouri for discussions.
\newline
B. Kyung$^{1}$ J. S. Landry,$^{1}$ D. Poulin$^{1}$ and A.-M. S. Tremblay$%
^{1,2}$\cite{email} \newline
$^{1}$D\'{e}partement de Physique and Centre de Recherche sur les propri\'{e}%
t\'{e}s \'{e}lectroniques de mat\'{e}riaux avanc\'{e}s\newline
$^{2}$Institut canadien de recherches avanc\'{e}es \newline
Universit\'{e} de Sherbrooke, Sherbrooke, Qu\'{e}bec, Canada J1K 2R1 \newline
\newline
Received 14 December 2001 \newline
DOI:  \newline
PACS numbers: 71.27.+a, 71.30.+h \newline

\end{document}